\documentclass[runningheads]{llncs}
\usepackage{hyperref}
\usepackage{amsmath}
\usepackage{cite}
\usepackage{algorithm}
\usepackage[noend]{algpseudocode}
\usepackage[section]{placeins}
\usepackage[bottom]{footmisc}
\usepackage{tikz}
\usepackage{blindtext}
\usepackage{xpatch}
\usepackage{lmodern}
\usepackage{multicol}
\usetikzlibrary{patterns}
\usepackage{graphicx}
\usepackage{caption}
\usepackage{url}
\usepackage[T1]{fontenc}
\usepackage{listings}
\usepackage{pgfplotstable}
\usepackage{pgfplots}
\pgfplotsset{compat=newest}
\usepackage{graphicx}
\usepackage{rotating}
\usepackage{tcolorbox}

\usepackage{bibnames}
\usepackage{booktabs}
\usepackage{lipsum}
\usepackage{subcaption}
\usepackage{amsmath,amsfonts,amssymb}
\usetikzlibrary{automata,positioning}
\abovecaptionskip
\belowcaptionskip

\usepackage{xcolor}
\usepackage{rotating}

\usepackage{listings}
\usepackage{color}
\definecolor{dkgreen}{rgb}{0,0.6,0}
\definecolor{gray}{rgb}{0.5,0.5,0.5}
\definecolor{mauve}{rgb}{0.58,0,0.82}
\definecolor{darkblue}{rgb}{0.0,0.0,0.6}
\definecolor{cyan}{rgb}{0.0,0.6,0.6}
\definecolor{mBlue}{HTML}{4285f4}
\definecolor{mRed}{HTML}{ea4335}
\definecolor{mGreen}{HTML}{34a853}
\definecolor{mYellow}{HTML}{fbbc04}
\definecolor{mLightBlue}{HTML}{6d9eeb}
\definecolor{mLightRed}{HTML}{e06666}
\definecolor{mLightGreen}{HTML}{93c47d}
\definecolor{mLightYellow}{HTML}{ffd966}
\definecolor{archtBlue}{HTML}{9fc5e8}
\definecolor{archtTeal}{HTML}{a4d2d7}
\definecolor{archtYellow}{HTML}{ffe599}
\definecolor{archtOrange}{HTML}{f9cb9c}
\definecolor{archtCetus}{HTML}{ea9999}
\definecolor{archtOther}{HTML}{dd7e6b}
\definecolor{archtPurple}{HTML}{b4a7d6}
\definecolor{archtGray}{HTML}{eeeeee}
\definecolor{archtGreen}{HTML}{b6d7a8}
\definecolor{archtRed}{HTML}{ea9999}

\lstdefinelanguage{XML}
{
  morestring=[b]",
  morestring=[s]{>}{<},
  morecomment=[s]{<?}{?>},
  stringstyle=\color{black},
  identifierstyle=\color{darkblue},
  keywordstyle=\color{cyan},
  morekeywords={xmlns,version,type,ma-id}
}

\newcolumntype{C}[1]{>{\centering\arraybackslash}m{#1}}
\makeatletter
\def\BState{\State\hskip-\ALG@thistlm}
\makeatother
\DeclareUrlCommand\UScore{\urlstyle{rm}}
\newcommand{\expUScore}{%
  \expandafter\expandafter\expandafter
  \UScore
  \expandafter\expandafter\expandafter
}

\usepackage[title]{appendix}
\usepackage{amsmath}
\usepackage{algorithm}
\usepackage[noend]{algpseudocode}
\usepackage[bottom]{footmisc}
\usepackage{tikz}
\usetikzlibrary{shapes, arrows}
\usetikzlibrary{matrix, positioning, fit}
\usepackage{caption}
\usepackage{subcaption}
\usepackage{tabularx}
    \newcolumntype{L}{>{\raggedright\arraybackslash}X}
\usepackage{csvsimple}
\usepackage{url}
\usepackage[T1]{fontenc}
\usepackage{listings}
\usepackage{pgfplotstable}
\usepackage{pgfplots}
\pgfplotsset{compat=newest}
\usepackage{graphicx}
\usepackage{rotating}
\usepackage[a4paper, total={6in, 8in}]{geometry}
\hypersetup{
    bookmarks=true,         
    unicode=false,          
    pdftoolbar=true,        
    pdfmenubar=true,        
    pdffitwindow=false,     
    pdftitle={Certificate},    
    pdfauthor={Idan Mosseri},     
    pdfsubject={\textit{ComPar}: Optimized Compiler for Automatic OpenMP Source-to-Source Parallelization using Code Segmentation and Hyperparameters Tuning},   
    pdfcreator={Gal Oren},   
    pdfproducer={},  
    pdfkeywords={},
    pdfnewwindow=true,      
    colorlinks=false,       
    linkcolor=red,          
    citecolor=green,        
    filecolor=magenta,      
    urlcolor=cyan           
}

\usepackage{geometry}
\DeclareUrlCommand\ULurl{
  
  }

\begin{document}

\title{\textit{ComPar}: Optimized Multi-Compiler for Automatic OpenMP S2S Parallelization}

\titlerunning{\textit{ComPar}: Optimized Multi-Compiler for Auto' OpenMP S2S Parallelization}

\author{
Idan Mosseri\inst{1,2}
\and Lee-or Alon\inst{1,3}
\and Re'em Harel\inst{3,4}
\and Gal Oren\inst{1,2}\thanks{Corresponding author}}

\institute{
Department of Computer Science, Ben-Gurion University of the Negev, P.O.B. 653, Be'er Sheva, Israel 
\and Department of Physics, Nuclear Research Center-Negev, P.O.B. 9001, Be'er-Sheva, Israel
\and Israel Atomic Energy Commission, P.O.B. 7061, Tel Aviv 61070, Israel\\
\and Department of Physics, Bar-Ilan University, IL52900, Ramat-Gan, Israel
\email{idanmos@post.bgu.ac.il, alonlee@post.bgu.ac.il, reemharel22@gmail.com,  orenw@post.bgu.ac.il}}

\maketitle              
\begin{abstract}
Parallelization schemes are essential in order to exploit the full benefits of multi-core architectures, which have become widespread in recent years. In shared-memory architectures, the most comprehensive parallelization API is OpenMP. However, the introduction of correct and optimal OpenMP parallelization to applications is not always a simple task, due to common parallel shared-memory management pitfalls, architecture heterogeneity and the current necessity for human expertise in order to comprehend many fine details and abstract correlations. To ease this process, many automatic parallelization compilers were created over the last decade.~\cite{harel2020source} tested several source-to-source compilers and concluded that each has its advantages and disadvantages and no compiler is superior to all other compilers in all tests. This indicates that a fusion of the compilers’ best outputs under the best hyper-parameters for the current hardware setups can yield greater speedups. To create such a fusion, one should execute a computationally intensive hyper-parameter sweep, in which the performance of each option is estimated and the best option is chosen. 
We created a novel parallelization source-to-source multi-compiler named \textit{ComPar}, which uses code segmentation-and-fusion with hyper-parameters tuning to achieve the best parallel code possible without any human intervention while maintaining the program's validity. In this paper we present \textit{ComPar} and analyze its results on NAS and PolyBench benchmarks. We conclude that although the resources \textit{ComPar} requires to produce parallel code are greater than other source-to-source parallelization compilers -- as it depends on the number of parameters the user wishes to consider, and their combinations -- \textit{ComPar} achieves superior performance overall compared to the serial code version and other tested parallelization compilers. \textit{ComPar} is publicly available at: \ULurl{https://github.com/Scientific-Computing-Lab-NRCN/compar}.

\end{abstract}

\section{Introduction}
Since the end of Dennard scaling \cite{dennard1974design} in 2005, there is a growing usage in multi-core architectures. These architectures can be found in a wide range of computers from wearable devices through smartphones and personal computers to high-performance computers \cite{blake2009survey}. Although these architectures can yield excellent performance in theory, in practice one should adjust his programming methods to work in parallel \cite{pacheco2011introduction}, i.e. to be executed by several processing units simultaneously. Furthermore, to fully exploit these architectures, one has to consider balancing the workload of the program between the processing units. Unfortunately, transforming a program from a sequential into a parallel one may be a very complicated and pricey task, especially when dealing with legacy codes \cite{feathers2004working}. This is due to the fact that in order to evolve a program to work in a parallel fashion, one must have a deep understanding of the code behavior and be very cautious not to change the inner logic of the program while attempting to utilize the benefits of the system. In a shared-memory setting, this is usually done via compiler optimizations and parallelization API such as OpenMP.

OpenMP~\cite{dagum1998openmp} is a pragma (compiler directive) oriented library for shared memory parallelization. The programmer can mark code segments by wrapping them with directives that instruct the compiler how to perform the parallelization. At run-time, each code segment is divided and executed concurrently on several threads. Note that the compiler might ignore the suggested directives. In this case, the segment that was wrapped by the ignored directive will not be executed in parallel. In addition to the directives, OpenMP offers a wide variety of run-time sub-routines and environment variables that can control the run-time specification and the fashion of the parallel execution. All of the above have an impact on the final performance of the parallel execution. To ease the burden of introducing such directives, several source-to-source (S2S) parallelization compilers that allow users to automatically parallelize their code \cite{prema2019study} -- prior to the machine-code compilation -- were invented. 

The automatic S2S parallelization compilers insert parallelization instructions in different fashions while preserving both the program's correctness and data coherence implied by its data dependencies. These compilers work as follows: The compilers parse the code into an Abstract Syntax Tree (AST) \cite{neamtiu2005understanding}; then, they find data dependencies by analyzing the generated tree; and afterward, they add parallel directives to certain code segments in an attempt to optimize the performance of the code. This process is done several times until convergence. At the end of the process, the tree is converted back to code in the original programming language.
The following note should be highlighted in this context: Currently, no existing automatic parallelization compiler can fully replace the programmer's insight, as programmers are still able to push the performance of the parallelization further than automatic compilers. This is since some information is usually hard to automatically extract from the AST alone, and is crucial for full exploitation of the parallel performance of the code. For example, function side effects; pointer aliasing; valuable information that may be based on computational load; optimal scheduling; chunk size and the number of threads. In this work, we introduce \textit{ComPar}: a unified multi-compiler that sweeps over different hyper-parameters for each code segment that is suitable for parallelization using automatic S2S parallelization compilers and fuses the best results, in terms of performance, together into one optimal code.

The rest of the paper is organized as follows: In Section~\ref{relatedwork} we present the related work done in regards to automatic parallelization compilers and the foundations of \textit{ComPar}. In Section~\ref{sec_chosen_compilers} we briefly discuss the relevant compilers for \textit{ComPar} purposes. In Section~\ref{sec_compar} we present \textit{ComPar}, and examine its performance in Section~\ref{sec_experiments}. Finally, we conclude this work and discuss future work in Section~\ref{sec_conclusions}.

\section{Related Work}\label{relatedwork}
\textbf{S2S Automatic Parallelization Compilers:} 
S. Prema et al.~\cite{prema2017identifying} compared several automatic parallelization compilers (not necessarily S2S) including Cetus \cite{dave2009cetus}, Par4All \cite{amini2012par4all}, Pluto \cite{bondhugula2007pluto}, Parallware \cite{parallware, gomez2015novel}, ROSE \cite{rosehome, quinlan2000rose}, and ICC \cite{icc}. They discussed the different aspects of the compilers' work fashions and showed their speedups and points of failure on ten NAS Parallel Benchmarks~\cite{bailey1991parallel} using the Gprof performance analysis tool~\cite{graham2004gprof}. While Parallware and Pluto failed to parallelize the benchmarks, the authors suggested a way to overcome these points of failure with manual intervention. They observed that Par4All requires no manual intervention, while Cetus and AutoPar require minimal manual intervention, thus allowing us to consider them for this work. Harel et al.~\cite{harel2020source} focused on Cetus, Par4All, and AutoPar \cite{autopardoc} while eliminating the need for the rest of the S2S automatic parallelization compilers. ~\cite{harel2020source} briefly discussed these compilers (regarding both history and work fashion) and presented each compiler's strengths and weaknesses. Moreover, ~\cite{harel2020source} tested the performance of these compilers in the Matrix Multiplication kernel and the NAS benchmark \cite{bailey1991parallel}. In addition,~\cite{harel2020source} pointed out the pitfalls of the selected compilers and proposed changes to their code-base, in an attempt to aid these compilers to insert more OpenMP directives.~\cite{harel2020source} also compared the compilers' performance on two different suitable hardware architectures -- multi-core (Non-Uniform Memory Access) and many-core (XeonPhi, GPGPU). ~\cite{harel2020source} concluded that currently there is no best S2S automatic parallelization compiler. However, there is a preferable compiler for each specific case, as the compilers behave differently either inherently (e.g. different AST analysis and precautions) or extrinsically (e.g. compilation flags of the parallelizer itself), thus finding the preferable one might be a tedious and costly task.

\textbf{Hyper-parameters Tuning:} 
The concept of auto-tuning OpenMP code is well-established \cite{katarzynski2014towards, liao2009effective, mustafa2011performance, silvano2018autotuning}, and as one can assume, the choice of each environment variable can greatly affect the performance of the code ~\cite{balaprakash2018autotuning}. Consider for example the $dynamic$ scheduling option: If the chosen $chunk\_size$ is too small, the resulting numerous work segments cause high overhead. Contrary, too large $chunk\_size$ may result in some threads that will not be assigned with any work, hence harming the parallelization performance. Therefore, these variables should be carefully tuned. One way to do this is by testing and empirically selecting the optimal ones. Sreenivasan et al.~\cite{sreenivasan2019framework} 
proposed an auto-tuning tool for OpenMP directives. The 
suggested framework currently supports only changing the number of threads used for parallel regions (the more the merrier does not necessarily apply here), the $chunk\_size$, and the scheduler type ($static$/$dynamic$). However, in addition to these control variables, recent advancements in OpenMP provides many additional variables that control the run-time environment of the program, which may increase the performance of the program when defined correctly \cite{van2017using}. For example, even in the context of the already used variables, \cite{sreenivasan2019framework} disregarded newer types of scheduling such as $guided$, $auto$, and $runtime$.

\textbf{Code Segmentation-and-Fusion:} 
As OpenMP directives target each optional parallel section separately (in contrast, for example, to MPI \cite{gropp1999using}), and as each one of them might have a completely different work fashion and balance, no unified compilation of an entire program using a single S2S compiler can assure the best possible performance. Thus, code segmentation into possibly parallel sections, followed by a varied S2S compilation sweep for best match in terms of performance is needed. Although not S2S, this idea was previously suggested by Shivam et al. in MCompiler~\cite{shivam2019mcompiler}, which divides the code into segments, chooses the best machine-code compiler for each segment, and composes the compiled segments back together. MCompiler uses the following compilers: Intel's C compiler \cite{icc}, PGI's C compiler \cite{pgi}, GNU GCC \cite{gnu}, LLVM Clang \cite{lattner2008llvm}, Polly \cite{grosser2011polly}, and Pluto \cite{bondhugula2007pluto}. MCompiler's code segmentation is based on identifying loops in the code. While compiling a loop nest, MCompiler attempts to optimize it using different compiler flags. Machine learning is optionally used in order to match each loop nest to the proper compiler before running the job in practice. However, the reliance of MCompiler on machine-code compilation to gain higher performance and not on S2S with an OpenMP parallelization, prevents users from retrieving the enhanced code for further development, as well as tweaking run-time variables such as the number of threads used by the computation or other parallelization-related ones. Yet, MCompiler may be used as the machine-code compiler for resulted S2S automatic parallelized code, thus achieving better performance both in terms of machine compilation as well as parallelization.

\textbf{Unified Multi-Compiler Approach:}
Concluding,~\cite{harel2020source} suggested an automatic compiler that will take the current automatic parallelization performances to the next level: Dividing the code into suitable-for-parallelization segments, choosing the best parallelization compiler for each segment while tuning the hyper-parameters (both OpenMP's and the compiler's) and fusing the outperforming segments back together to a unified code. The suggested compiler is based on the assumption that there is no best compiler for an \textit{entire} program, yet there is one for a suitable-for-parallelization individual \textit{segment}, as each compiler is preferable for a different task under different hyper-parameters. As High-Performance Computing (HPC) resources skyrocket over the last decade, such a compute-intensive task of hyper-parameters sweep and the execution of many computations to achieve the best performing code is no longer impossible in terms of computing power and might be worthwhile and cost-effective for long-living and legacy codes. Moreover, as those codes use HPC resources constantly and on a massive scale, even modest optimizations to the codes' performances -- in terms of parallelization efficiency -- can dramatically reduce future unnecessary stacked costs. Ergo, in this paper, we implemented and extended the suggested compiler, named \textit{ComPar}.

\section{\textit{ComPar}'s S2S Automatic Parallelization Compilers}\label{sec_chosen_compilers}
As ~\cite{harel2020source} concluded, AutoPar, Par4All, and Cetus are the most suitable compilers for S2S automatic parallelization (although other S2S compilers can be easily added to \textit{ComPar} by implementing an appropriate interface). Therefore, we decided to incorporate them into \textit{ComPar}. In the following section a brief summary of each chosen compiler is provided.

\textbf{Cetus:} Cetus \cite{cetushome} is an open-source S2S automatic parallelization compiler for C programs, which was developed by the ParaMount research group at Purdue University. Cetus compiler can verify existing OpenMP directives in a given code and perform data-dependent analysis, pointer alias analysis, and array privatization and reduction recognition. Moreover, Cetus uses a special flag to guarantee that parallelization is done only for loops above 10,000 iterations, in an attempt to prevent parallelization overhead. In cases of nested loops, the number of iterations of each loop segment will also include the number of iterations of its inner loops. However, standard compilers may not recognize Cetus' clauses. One main disadvantage of Cetus is that it does not insert OpenMP directives to loops that contain function calls. 

\textbf{AutoPar:} AutoPar \cite{autopardoc} is an open-source S2S automatic parallelization compiler for C and C++ programs and is developed by Lawrence Livermore National Laboratory (LLNL). Besides AutoPar's ability to automatically insert OpenMP directives to a given code, it can also ensure the correctness of the directives in a given parallel code. As was mentioned above, some additional manual information is required from the user in order to maximize the parallelization performance. Users can provide to AutoPar an annotation file describing the features of the code.

\textbf{Par4All:} Par4All \cite{par4allhome} is an open-source S2S automatic parallelization compiler for C and Fortran programs, which was developed by SILKAN, MINES ParisTech, and Institute T\'el\'ecom as a merge of some open-source development projects. This compiler is suitable for a broad range of hardware architectures~\cite{amini2012par4all}, and in particular it can be used to migrate programs to multi-core processors and GPGPUs using CUDA paradigms. Furthermore, it can optimize code execution on multi-core and many-core architectures. Par4All can perform data dependencies analysis and can validate the correctness of code manipulations. Note that Par4All may change the structure of the code.

\section{\textit{ComPar}: From Theory to Practice}\label{sec_compar}
As was discussed in~\cite{harel2020source}, each tested compiler has its advantages and disadvantages and no compiler is superior to the other compilers in all tested benchmarks. Hence, using only one compiler at a time is not enough in order to reach optimal performance. This might suggest that one should carefully fuse the abilities of all compilers in order to fully exploit the given hardware capabilities to the limit. In this paper, we suggest \textit{ComPar} - a novel parallelization S2S compiler that follows this vision.

\subsection{Characteristics, Architecture and Workflow}
\textit{ComPar} is a S2S compiler that optimizes the parallelization of the code in terms of running-time that can be achieved from S2S automatic parallelization compilers without any human intervention. This is done by fusing several outputs of said compilers while selecting the best from each based on varied empirical tests. \textit{ComPar} only requires the user to specify the desired hyper-parameters to be considered (i.e. the parameters defined by OpenMP and the different compilers) in a JSON format. Note that although, theoretically, \textit{ComPar} considers all available compilers' flags as well as OpenMP \textit{parallel for} directive clauses and OpenMP run-time library routines, some of them might affect the correctness of the program. The correctness of the generated code is based on the assumption that it is the responsibility of the user to provide reasonable \textit{guiding} parameters, as the user is familiar with the logic of the source code, its dependencies, and the hardware at hand. For example, in cases of a source code containing pointer aliasing, the user must not provide the \textit{no-pointer-aliasing} flag as a parameter in the JSON file. We suggest two methods to overcome this problem: (1) \textit{ComPar}'s black-box testing functionality, which examines the functionality of an application before and after the parallelization without peering into its internal structures or workings, and (2) AutoPar's ability to ensure the correctness of OpenMP directives in a given parallel code. 

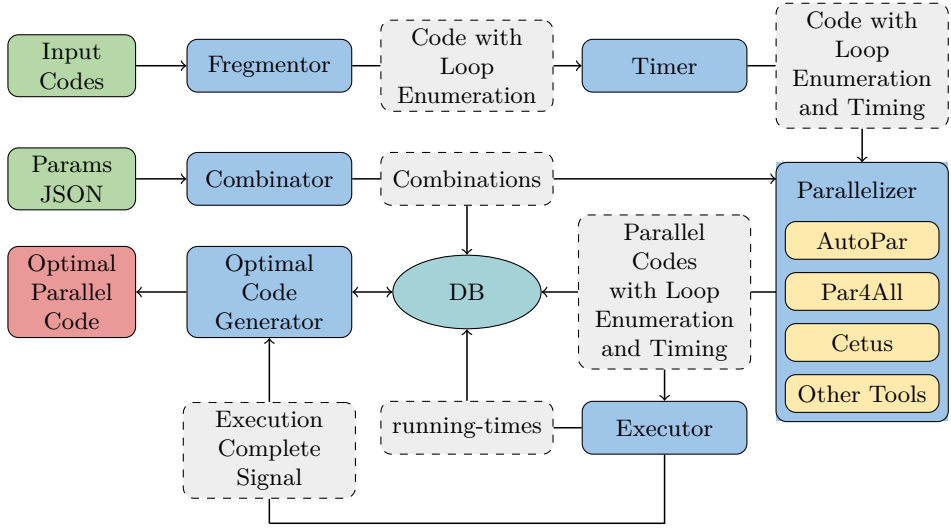
\begin{figure}[ht!]

  \centering
    
\tikzstyle{input} = [rectangle, draw, fill=archtGreen, minimum width=1cm, 
    text width=4.5em, text centered, rounded corners, minimum height=0.7cm]
    
\tikzstyle{output} = [rectangle, draw, fill=archtRed, minimum width=1cm, 
    text width=4.5em, text centered, rounded corners, minimum height=0.7cm]
    
\tikzstyle{middle_logic} = [rectangle, draw, fill=archtBlue, minimum width=1cm, 
    text width=6.0em, text centered, rounded corners, minimum height=0.7cm]

\tikzstyle{middle_parts} = [rectangle, draw, fill=archtGray, minimum width=1cm, 
    text width=6.3em, text centered, rounded corners, minimum height=0.7cm, dashed]

\tikzstyle{line} = [draw, -latex']

\tikzstyle{cloud} = [draw, ellipse,fill=archtTeal,
    minimum height=3em, minimum width=6em]
    
\begin{tikzpicture}[mymatrix/.style={matrix of nodes, nodes=typetag, fill=archtBlue, row sep=0.5em},
  mycontainer/.style={draw=black, inner sep=0ex, text centered, rounded corners},
  typetag/.style={draw=black, inner sep=0.5ex, anchor=west, text centered, rounded corners, text width=5.8em, fill=archtYellow!80, minimum height=0.5cm},
  title/.style={draw=none, color=black, fill=archtBlue, inner sep=0pt, text centered} ,node distance = 1.5cm, auto]
    
    \node [input] (input_codes) {Input\\Codes};
    
    \node [input, below of=input_codes] (json) {Params\\JSON};
    
    \node [output, below of=json] (output_code) {Optimal Parallel\\Code};
    
    
    \node [middle_logic, right of=input_codes, node distance=2.6cm] (fragmentor) {Fregmentor};
    
    \node [middle_logic, right of=json, node distance=2.6cm] (combinator) {Combinator};
    
    \node [middle_logic, right of=output_code, node distance=2.6cm] (generator) {Optimal Code Generator};
    
    \node [middle_parts, below of=generator, node distance=2.1cm] (execution_complete_signal) {Execution Complete Signal};
    
    
    \node [middle_parts, right of=fragmentor, node distance=2.6cm] (loop) {Code with Loop \\Enumeration};
    
    \node [middle_parts, below of=loop] (combinations) {Combinations};
    
    \node [cloud, below of=combinations] (db) {DB};

    \node [middle_parts, below of=db, node distance=1.8cm] (running-times) {running-times};
    
    
    \node [middle_logic, right of=loop, node distance=2.6cm] (timer) {Timer};
    
    \node [middle_parts, right of=db, node distance=2.6cm] (parallel) {Parallel Codes with Loop Enumeration and Timing};
    
    \node [middle_logic, below of=parallel, node distance=1.8cm] (executor) {Executor};
    
    
    \node [middle_parts, right of=timer, node distance=2.6cm] (loop_and_timing) {Code with Loop\\Enumeration\\and Timing};
    
  \matrix[mymatrix, below of=loop_and_timing, node distance=3cm] (mx1) {
    |[title]|Parallelizer \\
    AutoPar \\
    Par4All \\
    Cetus \\
    Other Tools \\
  };
  
  \node[mycontainer, fit=(mx1)] {};
    
    \draw [->, line width=0.2mm, black] (input_codes) -- (fragmentor);
    \draw [->, line width=0.2mm, black] (json) -- (combinator);
    \draw [<-, line width=0.2mm, black] (output_code) -- (generator);
    
    \draw [line width=0.2mm, black] (fragmentor) -- (loop);
    \draw [line width=0.2mm, black] (combinator) -- (combinations);
    \draw [<->, line width=0.2mm, black] (generator) -- (db);
    
    \draw [->, line width=0.2mm, black] (running-times) -- (db);
    
    \draw [->, line width=0.2mm, black] (combinations) -- (db);
    
    \draw [->, line width=0.2mm, black] (loop) -- (timer);
    
    \draw [line width=0.2mm, black] (timer) -- (loop_and_timing);
    
    \draw [->, line width=0.2mm, black] (loop_and_timing) -- (mx1);
    
    \draw [->, line width=0.2mm, black] (combinations.east) +(0,+0em) coordinate (b1) -- (mx1.west |- b1);
    
    \draw [->, line width=0.2mm, black] (parallel) -- (db);
    
     \draw [->, line width=0.2mm, black] (execution_complete_signal) -- (generator);
     
     \draw [->, line width=0.2mm, black] (parallel) -- (executor);
     
     \draw [line width=0.2mm, black] (executor) -- (running-times);
     
     \draw [line width=0.2mm, black] (execution_complete_signal.south)+(0,-1em) coordinate (a) -| (executor.south);
     \draw [line width=0.2mm, black] (execution_complete_signal.south)+(0,-1em) coordinate (a) -- (execution_complete_signal.south);
     
     \draw [line width=0.2mm, black] (mx1) -- (parallel);

\end{tikzpicture}
\vspace*{0.5mm}
   \caption{Architecture and workflow of \textit{ComPar}. Green: Inputs, Blue: Modules, Grey: Transferred data type, Yellow: Compilers, Teal: DB, Red: Output.}
  \label{fig:compar_archi}
  \end{figure}

The workflow of \textit{ComPar} is as follows (summarized in the diagram in Fig.~\ref{fig:compar_archi}):
First, the \textit{Fragmentor} enumerates and annotates all loops in the given source code. Next, the \textit{Timer} adds a piece of code around each enumerated loop which will later be used to measure its execution time. Meanwhile, the \textit{Combinator} parses three JSON files specifying which S2S compilers should be used; which compilation flags should be considered for each compiler; which OpenMP directives should \textit{ComPar} consider adding to each parallel loop (i.e. \textit{schedule(kind[, chunk\_size]})); and which OpenMP RTL functions should \textit{ComPar} consider adding before each loop. The \textit{Combinator} registers a combination in the \textit{DB} for each possible permutation of the above parameters. Consequently, For every such combination, the \textit{Parallelizer} parallelizes the code with the compiler and flags specified by the combination, and then adds the specified directive clauses and RTL functions to the loops that the compiler parallelized. Each parallel code is then executed by the \textit{Executor}, which logs its total running-time and the running-times of all of its loops in the \textit{DB}. Finally, after all combinations are executed, the \textit{Optimal Code Generator} chooses the parallelization scheme that produced the shortest running-time across all combinations for every individual loop and creates a parallel code version in which each loop is parallelized using its empiric optimal parallelization scheme. 

Additionally, as previously noted, the user may provide \textit{ComPar} with a testing script that verifies the correctness of each execution according to its output (i.e \textit{stdout} or output file). Using this script, \textit{ComPar} rejects any combination that did not pass the tests, thus providing correctness criteria that might help with pointing out invalid hyper-parameters. The user can also use \textit{AutoPar}'s abilities in this regard.

Assuming the correctness of the input, and the complete preservation of the entire AST under each S2S compiler, the theoretical proof of \textit{ComPar} optimization is straight-forward. The algorithm chooses the best directive provided by the different compilers for each loop segment. Thus, \textit{ComPar} either improves or does not change the running-time of the parallelized algorithm that could be produced by the best compiler, i.e., in the worst case, \textit{ComPar}'s output would be the best-parallelized code out of the codes that were generated by each of the supported compilers separately (or the serial code in case none of them succeed). We stress that a decrease, improvement or disruption of the code performance or results can be an outcome \textit{only} of the selected parallelization paradigm per each segment, and that the code validity can be assured using \textit{ComPar}'s black-box testing functionality and AutoPar's ability to ensure the correctness of the OpenMP directives in a given parallel code.

As was mentioned above, \textit{ComPar} runs all possible combinations of S2S compilers and flags, thus the number of combinations is given by the number of subsets of possible flags, which is:
$$
\sum_{i \in C}(2^{n_i} - 1)(2^{rtl + d} -1)
$$
where $C$ is the group of S2S compilers, $n_i$ is the number of flags to consider for S2S compiler $i$, and $rtl$ and $d$ are the number of run-time library routines and directives to consider adding to parallel loops, respectively.

The running-time of a single combination is the running-time of the corresponding parallel version of the input code, thus the total time until \textit{ComPar} produces its output is the sum over all the running-times of all combinations. Since \textit{ComPar}'s running-time depends on the running-time of the given source code, if one wishes to parallelize code $s$, it is strongly recommended to choose a sufficiently suitable input $x'$ for $s$, preferably a 'sweet-spot' in which the input is not too small to cause the parallel code to overwhelmingly suffer from parallelization overhead and not too big to cause the code to suffer from excessive running-times. Only then it is recommended to run the realistic input $x$ using the parallel code generated. 

\subsection{Interface}
\begin{figure}
    \centering
    \begin{subfigure}[b]{\textwidth}
        \centering
        \includegraphics[width=0.9\textwidth]{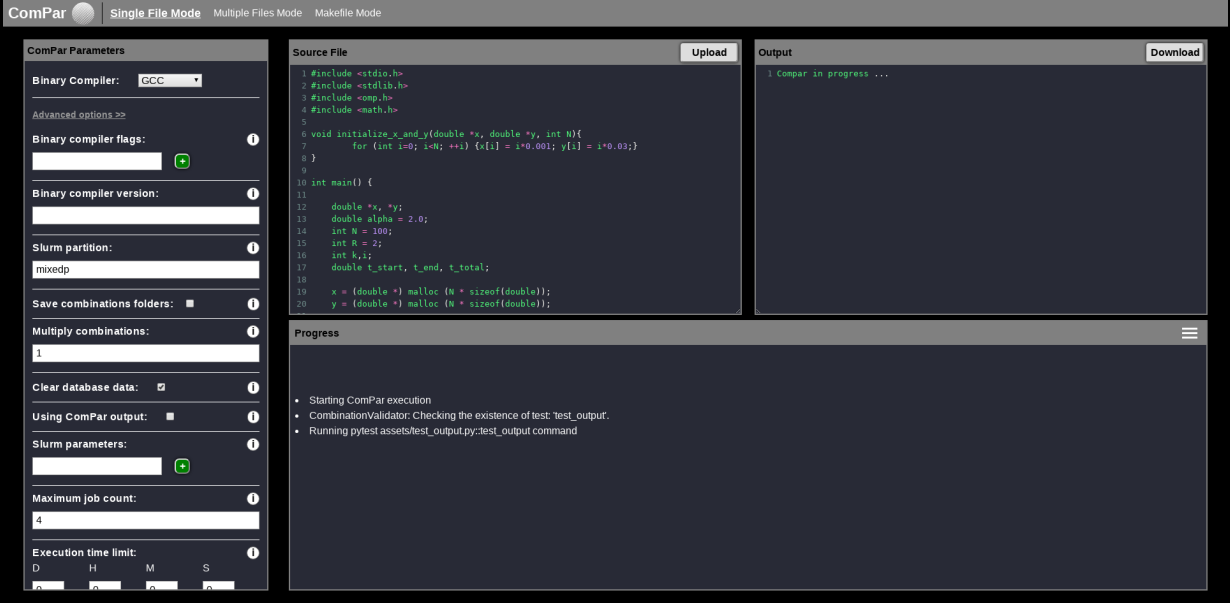}
        \caption{Single file mode}
        \label{fig:single_file_mode}
    \end{subfigure}
     
    \begin{subfigure}[b]{\textwidth}
        \centering
        \includegraphics[width=0.9\textwidth]{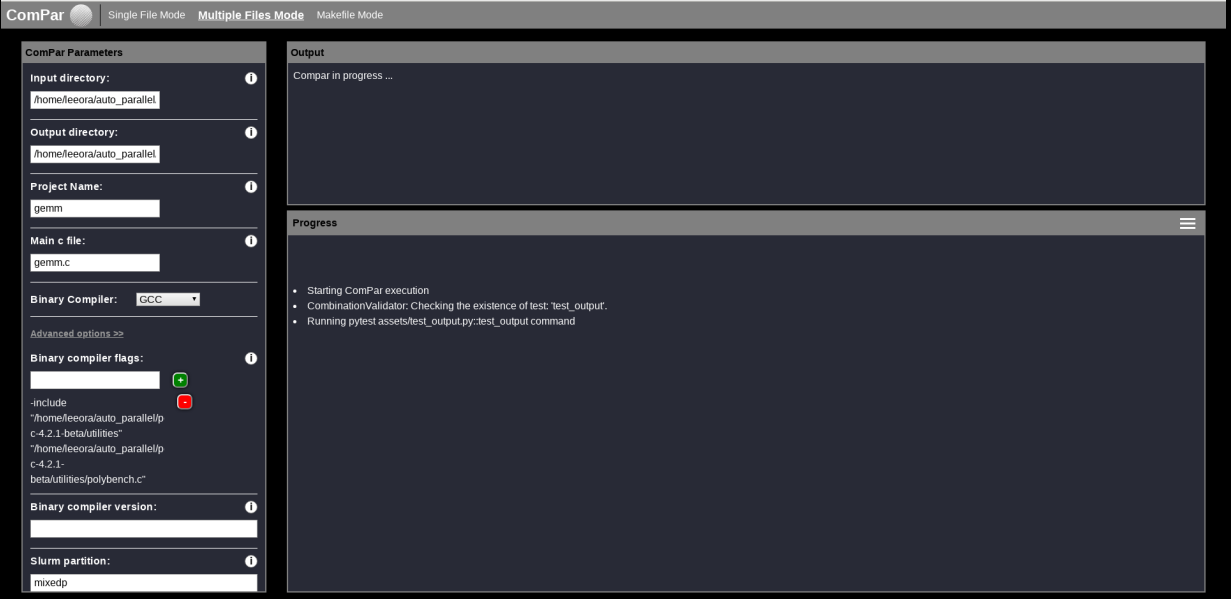}
        \caption{Multiple files mode}
        \label{fig:multi_files_mode}
    \end{subfigure}
    
    \begin{subfigure}[b]{\textwidth}
        \centering
        \includegraphics[width=0.9\textwidth]{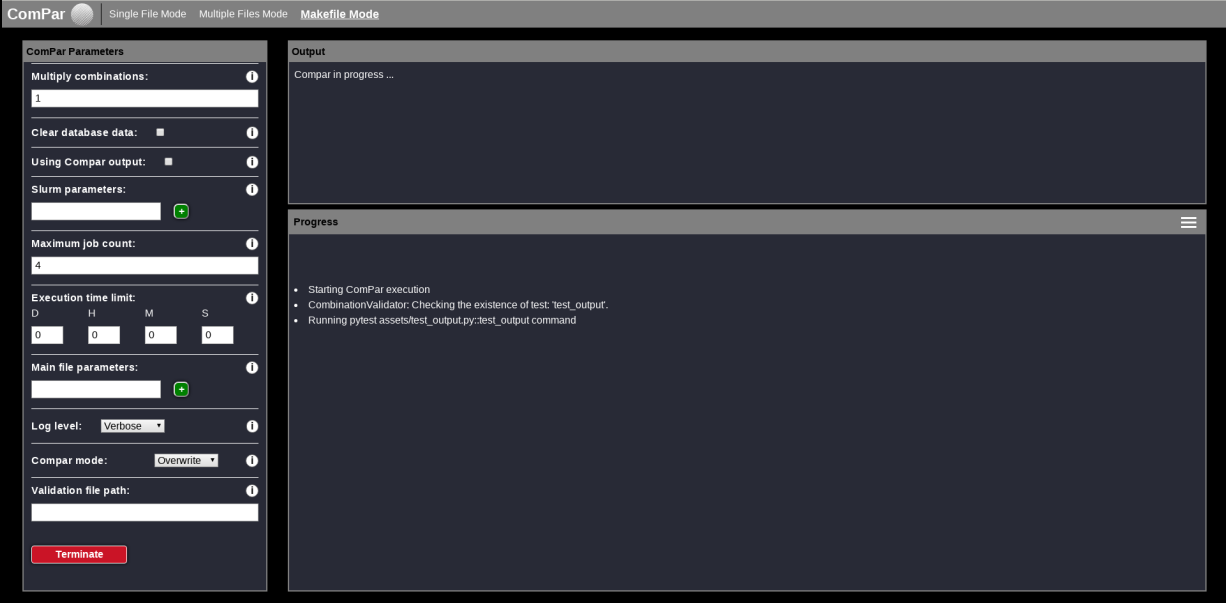}
        \caption{Makefile mode}
        \label{fig:makefile_mode}
    \end{subfigure}
    \vspace{0.15cm}
    \caption{ComPar's GUI three compilation modes: (1) Single file mode, used when there is only a single input source file, (2) Multiple file mode, used to process projects containing multiple files without a Makefile, and (3) Makefile mode, used to process Makefile projects.}
    \vspace{-5cm}
    \label{fig:gui}
\end{figure}

\textit{ComPar} offers both command-line and GUI interfaces with a verity of options such as compilation options, i.e. whether to use a Makefile or what machine-code compiler (e.g. GCC, ICC, etc.) to use, together with the corresponding compilation flags; SLURM parameters (\textit{ComPar} executes its jobs using the SLURM resource manager~\cite{slurm}); whether or not to save all the created combinations' files; where to store \textit{ComPar}'s output; what is the name of the project and what operational mode to use, etc. \textit{ComPar}'s three operational modes are:
\begin{enumerate}
      \item \textit{New}: This operational mode is used for new \textit{ComPar} executions. If a project with the same name already exists in \textit{ComPar}'s \textit{DB} under the same user, \textit{ComPar} will append an incremental index to the project's name, thus not overrunning previous executions. 
      \item \textit{Overwrite}: In this operational mode, previous executions of a project with the same name will be deleted and overwritten. 
      \item \textit{Continue}: This operational mode allows the user to resume a previous \textit{ComPar} exertion. this mode can be used to add more combinations or to resume a \textit{ComPar} execution that has crashed without re-running combinations that were already executed (on the same project). 
\end{enumerate}
The following are \textit{ComPar}'s GUI modes:
\begin{enumerate}
    \item \textit{Single File}: This mode is used when there is only a single input source file that the user wishes to parallelize (App. A, Fig.~\ref{fig:single_file_mode}).
    \item \textit{Multiple Files}: This mode is used to process projects that contain multiple files that do not have a Makefile (App. A, Fig.~\ref{fig:multi_files_mode}).
    \item \textit{Makefile}: This mode is used to process Makefile projects (App. A, Fig.~\ref{fig:makefile_mode}).
\end{enumerate}
The Single file mode layout is composed of four sections:
\begin{enumerate}
    \item \textit{Parameters}: In this area the user can view and edit all of \textit{ComPar}'s options mentioned above that are relevant to the current compilation mode.
    \item \textit{Source File}: In this area the user may upload/develop a single source file (relevant only to the Single File mode).
    \item \textit{Output}: The resulting parallel source file will be shown in this area at the end of \textit{ComPar}'s execution.
    \item \textit{Progress}: \textit{ComPar}'s output log is shown in this area.
\end{enumerate}
The full list of options can be found in \cite{compar_readme}.

\section{Experiments \& Discussion}\label{sec_experiments}
In order to evaluate the contributions of this paper, we examined the parallelization output on different kernels of both the NAS \cite{bailey1991parallel} and PolyBench ~\cite{polybench} Parallel benchmarks. \textit{ComPar} was compared against the different parallelization compilers and to serial executions. All of our benchmarks were executed using a single computation node with a total of 32 cores (AMD Opteron Processor 6376 \cite{amd}). Note that the number of threads utilized by the benchmark (correlates to the number of cores used) depends on each and every specification of combination. Table~\ref{compar_combination_params} presents the flags of the S2S compilers; the OpenMP \textit{parallel for} directive clauses; and OpenMP run-time library routines that we tested in our experiments. Moreover, we present the resulted speedups as well as the running-time in order to ratify the truthfulness of our results (by showing that they consumed a reasonable amount of computation time in regard to the given input and hardware settings).

\begin{table}[H]
\centering
    \begin{tabular}{|m{0.35\textwidth}|m{0.65\textwidth}|}
        \hline\cline{1-2}
        \multicolumn{2}{|l|}{\textbf{Compilers' Flags}} \\ 
        \hline\cline{1-2}
        \textit{Compiler} & \textit{Flag} \\
        \hline
            Cetus & parallelize-loops, reduction, privatize, alias \\
        \hline
            AutoPar & keep\_going, enable\_modeling, no\_aliasing, unique\_indirect\_index \\
        \hline
            Par4All & O, fine-grain, com-optimization, no-pointer-aliasing \\
        \hline\cline{1-2}
        \multicolumn{2}{|l|}{\textbf{OMP \textit{parallel for} Directive Clauses}} \\
        \hline\cline{1-2}
            \textit{Clause} & \textit{Kind} \\
        \hline
            schedule & static [2, 4, 8, 16, 32], dynamic \\
        \hline\cline{1-2}
        \multicolumn{2}{|l|}{\textbf{Runtime Library Routines}} \\
        \hline\cline{1-2}
            \textit{RTL Routine} & \textit{Argument} \\
        \hline
            omp\_set\_num\_threads & 2, 4, 8, 16, 32 \\
        \hline
    \end{tabular}
    \caption{Parallelization compilers' flags, OpenMP \textit{parallel for} directive clauses and OpenMP run-time library routines we tested in our experiments.}
    \label{compar_combination_params}
\end{table}
\vspace{-0.4cm}

\subsection{NAS Parallel Benchmarks}
\begin{figure}[H]
    \centering
    \begin{subfigure}[H]{\textwidth}
    \centering
    \makebox[0pt]{
    \begin{tikzpicture}
    \begin{axis}[
        width=\textwidth,
  	    height=5cm,
  	    ymode=log,
  	    log ticks with fixed point,
        ymajorgrids=true,
        ylabel={Running-time (sec)},
        enlarge x limits={abs=1.0cm},
        ybar,
        bar width=.24cm,
        xtick=data,
        xticklabels from table={compar_results/nas_runtime.csv}{Benchmark},
        log ticks with fixed point,
        every axis legend/.append style={
            at={(0.985,0.945)},
            anchor = north east,
            legend columns=3,
            legend cell align = right,
            draw = black,
        }
    ]
    \addlegendimage{fill=violet}
    \addlegendimage{fill=mBlue}
    \addlegendimage{fill=mRed}
    \addlegendimage{fill=mYellow}
    \addlegendimage{fill=mGreen}
    \addplot [fill=violet] table [x expr=\coordindex, y=Serial] {compar_results/nas_runtime.csv};
    \addplot [fill=mBlue] table [x expr=\coordindex, y=AutoPar] {compar_results/nas_runtime.csv};
    \addplot [fill=mRed] table [x expr=\coordindex, y=Par4All] {compar_results/nas_runtime.csv};
    \addplot [fill=mYellow] table [x expr=\coordindex, y=Cetus] {compar_results/nas_runtime.csv};
    \addplot [fill=mGreen] table [x expr=\coordindex, y=ComPar] {compar_results/nas_runtime.csv};
    \legend {
        Serial,
        AutoPar,
        Par4All,
        Cetus,
        ComPar
    }
    \end{axis}
    \end{tikzpicture}}
    \end{subfigure}
    \caption{NAS benchmark running-times achieved by the different compilers in logarithmic scale.}
    \label{nas_runtime_fig}
\end{figure}
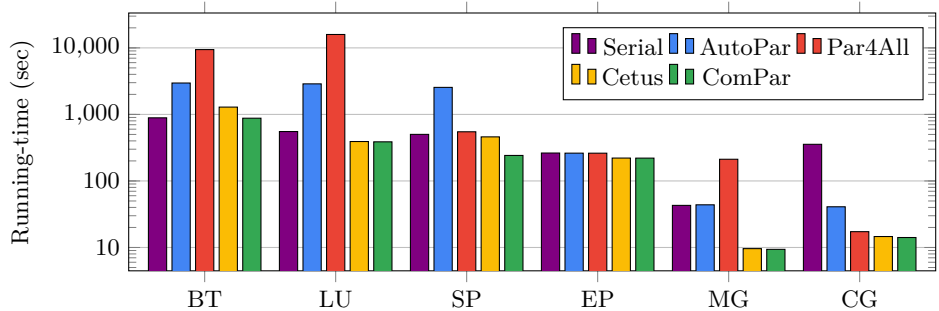

The Numerical Aerodynamics Simulations (NAS) Parallel Benchmarks \cite{bailey1991parallel} are a group of applications, developed by NASA, to evaluate the performance of high-performance computers. NAS Parallel Benchmarks include ten different benchmarks \cite{bailey2011parallel}. In order to be consistent with~\cite{harel2020source}, we tested the performance of the compilers over the following benchmarks: Block Tri-diagonal solver (BT), Conjugate Gradient (CG), Embarrassingly Parallel (EP), Lower-Upper Gauss-Seidel solver (LU), Multi-Grid (MG) and Scalar Penta-diagonal solver (SP). Similarly to~\cite{harel2020source}, we did not use Fourier Transform (FT), Integer Sort (IS) and Unstructured Adaptive mesh (UA) benchmarks, as some compilers failed to process them. As can be observed from Fig.~\ref{nas_speedup_fig}, \textit{ComPar} always achieved the best speedups, or at least the same ones as the best S2S compiler (which is different for each benchmark) (Fig.~\ref{nas_runtime_fig}).

\begin{figure}[H]
    \centering
    \begin{subfigure}[H]{\textwidth}
    \centering
    \makebox[0pt]{
    \begin{tikzpicture}
    \begin{axis}[
  	    width=\textwidth,
  	    height=5cm,
  	    ymode=log,
        ymajorgrids=true,
        ylabel={Speedup},
        enlarge x limits={abs=1cm},
        ybar,
        bar width=.24cm,
        xtick=data,
        xticklabels from table={compar_results/nas_speedup.csv}{Benchmark},
        log ticks with fixed point,
        every axis legend/.append style={
            at={(0.47,0.945)},
            anchor = north east,
            legend columns=3,
            legend cell align = right,
            draw = black,
        }
    ]
    \addlegendimage{fill=mBlue}
    \addlegendimage{fill=mRed}
    \addlegendimage{fill=mYellow}
    \addlegendimage{fill=mGreen}
    \addplot [fill=mBlue] table [x expr=\coordindex, y=AutoPar] {compar_results/nas_speedup.csv};
    \addplot [fill=mRed] table [x expr=\coordindex, y=Par4All] {compar_results/nas_speedup.csv};
    \addplot [fill=mYellow] table [x expr=\coordindex, y=Cetus] {compar_results/nas_speedup.csv};
    \addplot [fill=mGreen] table [x expr=\coordindex, y=ComPar] {compar_results/nas_speedup.csv};
    \legend {
        AutoPar,
        Par4All,
        Cetus,
        ComPar
    }
    \end{axis}
    \end{tikzpicture}}
    \end{subfigure}
    \caption{NAS benchmark speedups (compared to a serial execution) achieved by the different compilers in logarithmic scale.}
    \label{nas_speedup_fig}
\end{figure}
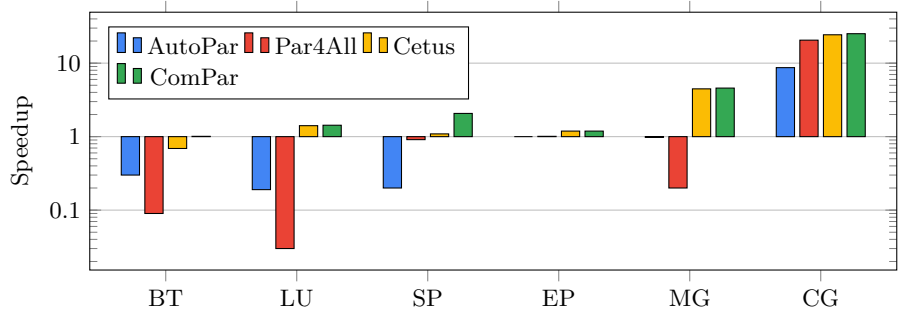

\vspace{-0.2cm}

\subsection{PolyBench Benchmarks}
PolyBench~\cite{polybench, pouchet2012polybench, yuki2014understanding} is a collection of 30 representative potentially compute-intensive benchmarks. It attempts to make the kernels' execution as uniform and consistent as possible. PolyBench contains a single file, tunable at compile-time, which is used for the kernel instrumentation. This file performs extra operations such as cache flushing before the kernels' execution, and can set real-time scheduling to prevent operating-system interference.

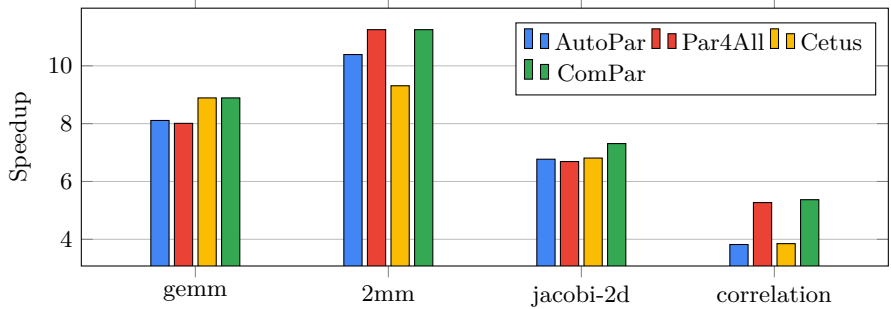
\begin{figure}[H]
\centering
    \begin{subfigure}[H]{\textwidth}
    \centering
    \makebox[0pt]{
    \begin{tikzpicture}
    \begin{axis}[
        width=\textwidth,
  	    height=5cm,
        ymajorgrids=true,
        ylabel={Speedup},
        enlarge x limits={abs=1.5cm},
        ybar,
        bar width=.24cm,
        xtick=data,
        xticklabels from table={compar_results/poly_speedup.csv}{Benchmark},
        log ticks with fixed point,
        every axis legend/.append style={
            at={(0.985,0.945)},
            anchor = north east,
            legend columns=3,
            legend cell align = right,
            draw = black,
        }
    ]
    \addlegendimage{fill=mBlue}
    \addlegendimage{fill=mRed}
    \addlegendimage{fill=mYellow}
    \addlegendimage{fill=mGreen}
    \addplot [fill=mBlue] table [x expr=\coordindex, y=AutoPar] {compar_results/poly_speedup.csv};
    \addplot [fill=mRed] table [x expr=\coordindex, y=Par4All] {compar_results/poly_speedup.csv};
    \addplot [fill=mYellow] table [x expr=\coordindex, y=Cetus] {compar_results/poly_speedup.csv};
    \addplot [fill=mGreen] table [x expr=\coordindex, y=ComPar] {compar_results/poly_speedup.csv};
    \legend {
        AutoPar,
        Par4All,
        Cetus,
        ComPar
    }
    \end{axis}
    \end{tikzpicture}}
    \end{subfigure}
    \caption{Polybench benchmark speedups (compared to a serial execution) achieved by the different compilers.}
    \label{poly_speedup_fig}
\end{figure}

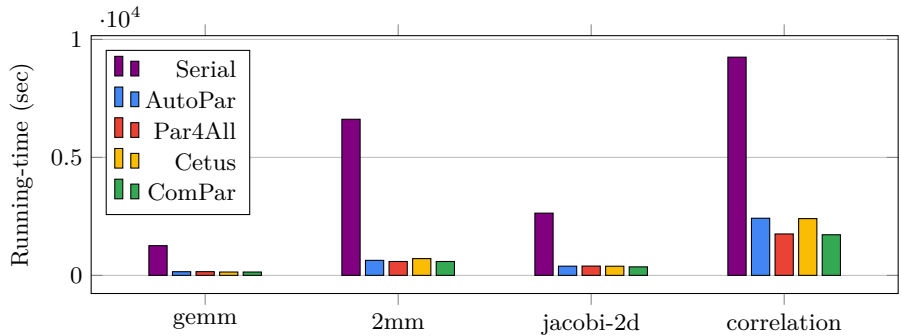
\begin{figure}[H]
\centering
    \begin{subfigure}[H]{\textwidth}
    \centering
    \makebox[0pt]{
    \begin{tikzpicture}
    \begin{axis}[
  	    width=\textwidth,
  	    height=5cm,
        ymajorgrids=true,
        ylabel={Running-time (sec)},
        enlarge x limits={abs=1.5cm},
        ybar,
        bar width=.24cm,
        xtick=data,
        xticklabels from table={compar_results/poly_runtime.csv}{Benchmark},
        log ticks with fixed point,
        every axis legend/.append style={
            at={(0.018,0.945)},
            anchor = north west,
            legend columns=1,
            legend cell align = right,
            draw = black,
        }
    ]
    \addlegendimage{fill=violet}
    \addlegendimage{fill=mBlue}
    \addlegendimage{fill=mRed}
    \addlegendimage{fill=mYellow}
    \addlegendimage{fill=mGreen}
    \addplot [fill=violet] table [x expr=\coordindex, y=Serial] {compar_results/poly_runtime.csv};
    \addplot [fill=mBlue] table [x expr=\coordindex, y=AutoPar] {compar_results/poly_runtime.csv};
    \addplot [fill=mRed] table [x expr=\coordindex, y=Par4All] {compar_results/poly_runtime.csv};
    \addplot [fill=mYellow] table [x expr=\coordindex, y=Cetus] {compar_results/poly_runtime.csv};
    \addplot [fill=mGreen] table [x expr=\coordindex, y=ComPar] {compar_results/poly_runtime.csv};
    \legend {
        Serial,
        AutoPar,
        Par4All,
        Cetus,
        ComPar
    }
    \end{axis}
    \end{tikzpicture}}
    \end{subfigure}
    \caption{Polybench benchmark running-times achieved by the different compilers.}
    \label{poly_runtime_fig}
\end{figure}

Most of the benchmarks in the same category are computationally comparable (e.g 2mm versus 3mm). Therefore, we chose one representative benchmark in each category (except for \textit{Medley} and \textit{Linear Algebra Solvers} which we considered redundant and highly time consuming in this context, respectively). We tested the performance of the compilers over \textit{correlation} (cat. Data Mining), \textit{gemm} (cat. BLAS), \textit{2mm} (cat. Linear Algebra Kernels), and \textit{jacobi-2d} (cat. Stencils). We did not change the number of iterations in any of the chosen benchmarks. However, we evenly enlarged the (already \textit{LARGE}) problem size by x8 (in terms of memory footprint) in order to ensure that the benefit from load-balancing imposed by the parallelization will not be overshadowed by the parallelization overhead.
Another benefit of maximizing memory usage (in regard to the given hardware) is that the running-time is less affected by the Non-Uniform Memory Access architecture and by the cache hierarchy, thus attempting to represent a full-scale job as much as possible. Again, as can be observed from Fig.~\ref{poly_speedup_fig}, \textit{ComPar} always achieved the best speedups, or at least the same ones as the best S2S compiler (which is different for each benchmark) (Fig.~\ref{poly_runtime_fig}).
\vspace{-0.2cm}

\section{Conclusions \& Future Work} \label{sec_conclusions}
In this paper, we address the pitfalls of S2S automatic parallelization and how some crucial aspects of them could be resolved using \textit{ComPar}. We briefly discussed Cetus, AutoPar and Par4All, which we found most suitable for this task. We then presented \textit{ComPar} and analyzed its results over both the NAS and the PolyBench benchmarks. We conclude that although the resources \textit{ComPar} consumes in order to produce efficient parallel code are greater than the resources other parallelization compilers demand -- as it depends on the number of parameters the user wishes \textit{ComPar} to consider -- \textit{ComPar} achieves superior overall performance compared to the tested parallelization compilers and the serial code version. We presented the reasons for which this usage might be worthwhile and even cost-effective.

Much work is left for the future: Adding support for Fortran programming language is one of our next goals, as \textit{ComPar} is primarily targeting legacy large-scale serial scientific codes. One may also try to better learn the code dependencies and refine the semantically correct parallelization parameters accordingly. Moreover, a comprehensive understanding of the hardware specs, let alone actively learning which hyper-parameters best suite each hardware using machine learning paradigms, may further enhance our speedups and shorten \textit{ComPar}'s execution time \cite{tournavitis2009towards}. In addition, the chosen S2S compilers are currently limited to OpenMP v2.5, hence the generated code can not utilize most of the advantages of directives from later OpenMP versions. Adding more automatic parallelization compilers might be also beneficial. Furthermore, adding more machine-code compilers might improve the current results and support additional input source codes. Currently, \textit{ComPar} can choose the most suited compiler for different hardware architectures only under certain circumstances (see section~\ref{sec_compar}), while in the future we wish to explore this improvement opportunity under other circumstances. As was discussed in~\cite{shivam2019mcompiler}, it may be advantageous to use VTune \cite{reinders2005vtune} in \textit{ComPar} in order to find the most suited automatic parallelization compiler for each code segment and the best machine-code compilers for each output file generated by \textit{ComPar} and each hardware architecture. Nevertheless, we emphasize that \textit{ComPar} is the first open-sourced platform for such optimizations of S2S automatic parallelization compilers, and as such could benefit from further unexplored avenues and future research. \newline

\textbf{Acknowledgments:} This work was supported by the Lynn and William Frankel Center for Computer Science. Computational support was provided by the NegevHPC project \cite{negevhpc}. The authors would like to thank Reuven Regev Farag, Gilad Guralnik, Yoni Cohen, May Hagbi, Shlomi Tofahi, and Yoel Vaizman from the Department of Software Engineering, Sami Shamoon College of Engineering, for their part in the development of \textit{ComPar}.

\bibliographystyle{unsrt} 
\bibliography{bibliography} 

\newpage
\section*{Appendix A: \textit{ComPar}'s GUI}
\label{sec_appendix}

\end{document}